\documentclass[aps,pra,reprint,twocolumn,groupedaddress,longbibliography,floats,final]{revtex4-2}
\usepackage[usenames,dvipsnames]{xcolor}
\usepackage{graphicx}
\usepackage{amssymb}
\usepackage{subfigure}
\usepackage{stackrel}
\usepackage{amsmath}
\usepackage{amsfonts}
\usepackage{bm}
\usepackage{graphicx}
\usepackage[normalem]{ulem}
\usepackage{longtable}

\definecolor{rosita}{rgb}{0.97, 0.56, 0.65}
\usepackage{amsmath}
\newcommand{\beq}{\begin{equation}}
\newcommand{\eeq}{\end{equation}}
\newcommand{\ket}[1]{\ensuremath{\left|{#1}\right\rangle}}

\newcommand{\braket}[2]{\ensuremath{\langle{#1}|{#2}\rangle}}
\newcommand{\op}[1]{\ensuremath{\hat{\mathnormal{#1}}}}
\newcommand{\mean}[1]{\ensuremath{\langle{#1}\rangle}}

\begin{document}

\title{The Underlying Order Induced by Orthogonality and the Quantum Speed Limit}

\author{Francisco J. \surname{Sevilla}}
\affiliation{Instituto de F\'{\i}sica, Universidad Nacional Aut\'{o}noma de M\'{e}xico, Apartado Postal 20-364, Ciudad de M\'{e}xico, Mexico.}
\email[]{fjsevilla@fisica.unam.mx}
\thanks{}

\author{Andrea Vald\'es-Hern\'andez}
\affiliation{Instituto de F\'{\i}sica, Universidad Nacional Aut\'{o}noma de M\'{e}xico,
Apartado Postal 20-364, Ciudad de M\'{e}xico, Mexico.}

\author{Alan J. \surname{Barrios}}
\affiliation{Instituto de F\'{\i}sica, Universidad Nacional Aut\'{o}noma de M\'{e}xico,
Apartado Postal 20-364, Ciudad de M\'{e}xico, Mexico.}

\begin{abstract}
We perform a comprehensive analysis of the set of parameters $\{r_{i}\}$ that provide the energy distribution of pure qutrits that evolve towards a distinguishable state at a finite time $\tau$,
when evolving under an arbitrary and time-independent Hamiltonian. The orthogonality condition is exactly solved, revealing a non-trivial interrelation between $\tau$ and the energy spectrum    and allowing the classification of $\{r_{i}\}$ into families organized in a 2-simplex, $\delta^{2}$. Furthermore, the states determined by $\{r_{i}\}$ are likewise analyzed according to their quantum-speed limit. Namely, we construct a map that distinguishes those $r_{i}$s in $\delta^{2}$ correspondent to states whose orthogonality time is limited by the Mandelstam--Tamm bound  from those restricted by the Margolus--Levitin one. Our results offer a complete characterization of the physical quantities that become relevant in both   the preparation and study of the dynamics of three-level states evolving towards orthogonality.
\end{abstract}
\maketitle

\section{Introduction}

The progress on the experimental manipulation of quantum states and control of quantum dynamics to achieve specific tasks  has been based on well-established theoretical elements that have pointed to new applications in quantum information science. One of these elements refers to the time $\tau$ required by an initial pure state to reach (for the first time) an orthogonal, distinguishable  state when evolved under a unitary (Hamiltonian) transformation \cite{FondaRepProgPhys1978,GiovannettiPRA2003,GislasonPRA1985,JonesPRA2010,JordanPRA2017,LevitinPRL2009,LuoPhysD2004,MarviaPRA2016,PfeiferRMP1995,
PiresPhysRevX2016,ZhangSciRep2014,ZielinskiPRA2006}. This {\it{orthogonality time}} $\tau$ represents the time required to perform an elementary computational step,   and   it establishes a characteristic scale for the dynamical evolution of the system. Its study has therefore both practical and fundamental \mbox{implications \cite{MirkinPRA2020,ShaoPRR2020,SuzukiPRR2020}}. 

Specially relevant is the \emph{minimal} amount of time required to reach orthogonality, setting a lower bound for $\tau$ known as the \emph{quantum speed limit} (QSL), $\tau_{\textrm{qsl}}$. Fundamental limitations on the QSL have been discovered \cite{Mandelstam1991,GislasonPRA1985,VaidmanAJP1992, PfeiferPRL1993,BraunsteinAnnPhys1996}, particularly related to the state's energy dispersion in the form of the Mandelstam--Tamm (MT) bound    or    to the state's mean energy (relative to the minimum energy), as discovered by Margolus and Levitin \mbox{(ML) \cite{MargolusPhysicaD1998}.} Since then, there has been an impressive progress on the investigations regarding the minimal amount of transformation linking two orthogonal states (see, e.g., the reviews \cite{FreyQInfProcc2016,DeffnerJPhysA2017}).

In their seminal work,  Levitin and Toffoli \cite{LevitinPRL2009} investigated the tightness of the unified bound that encompasses the MT and the ML bound.   They  demonstrated that only equally-weighted superpositions of two energy eigenstates saturate both bounds \emph{and} attain the QSL, whereas for superpositions of three energy eigenstates the unified bound is tight, so $\tau$ may be arbitrarily close, although not equal, to $\tau_{\text{qsl}}$. Further investigations consider more general situations as the case of systems of qubits \cite{ZhangPLA2018} or mixed quantum states \cite{TaddeiPRL2013,CampaioliPRL2018}. More recently, an experimental confirmation of the unified bound has been observed using fast matter wave interferometry of a single atom moving in an optical trap \cite{Ness2021},   and    a theoretical analysis of the feasibility for measuring QSLs in ultracold atom experiments  was  presented  by  \citet {delCampoPRL2021}.

Despite the progress boosted by Levitin and Toffoli, a thorough analysis of the necessary and sufficient conditions for a pure state, in a low-dimensional Hilbert space, 
to evolve towards an orthogonal one in a finite $\tau$ is far from having being completed and is still a pending task. The present paper contributes to this effort by presenting a detailed and comprehensive study of the conditions on both: the expansion coefficients $\{r_i\}$ of the state in the energy representation, and the system's accessible energy levels, that guarantee that a given initial (pure) state in a three-dimensional Hilbert space attains a distinguishable one in a finite amount of time.
 This  analysis is presented in Section \ref{sect:Orthogonality} for states that evolve under the action of a time-independent, otherwise arbitrary, Hamiltonian. Starting from the orthogonality condition, we classify the allowed sets $\{r_1,r_2,r_3\}$
  into two main families and establish their precise relation with the energy-level spacing and $\tau$. The relation among the coefficients, the energy-level spacing,  and the orthogonality time is afterwards represented in a diagram in which the families are depicted. 
Furthermore, the set of allowed $r_i$s is recognized as a 2-simplex contained in the probability 2-simplex of $\mathbb{R}^{3}$. We further investigate, in Section \ref{sect:QSL}, whether the quantum speed limit of the states determined by the coefficients $\{r_i\}$ found  is given by either   the MT or the ML bound. Finally,  we present a summary and some concluding remarks in Section \ref{sect:Conclusions}.

Our results disclose the exact and non-trivial interrelation among  the orthogonality time, the Hamiltonian eigenvalues    and    the parameters that provide the energy distribution in qutrit systems, which have potential applications in the dynamics of neutrino \mbox{oscillations \cite{KhanQInfProc2021},}   the laser-driven transformations of an atomic qutrit \cite{VitanovPRA2012},   the quantum Fourier transform of a superconducting qutrit \cite{YurtalanPRL2020}    or      the transfer of quantum information \cite{DelgadoPLA2007}. They also offer a complete characterization of practical value when attempting to prepare low-dimensional states that evolve towards a distinguishable one, either 
by specifying the appropriate initial state once a specific Hamiltonian is given or an appropriate transformation whenever the initial state is fixed. 

\section{\label{sect:Orthogonality} Necessary and Sufficient Conditions for Reaching Orthogonality}
We start   by carrying  out an analysis in the three-dimensional Hilbert space $\mathcal{H}_{3}$, to determine the conditions that drive a pure state under an arbitrary (time-independent) Hamiltonian evolution into an orthogonal state in a finite time. Our analysis is general enough and does not depend on the peculiarities of the physical system, opening the possibility of analyzing the minimal amount of transformation between orthogonal states for specific transformations of interest in the engineering of quantum computation. 

An arbitrary initial pure state $\ket{\psi(0)}$ in $\mathcal{H}_{3}$ can be expanded in the eigenbasis $\{\ket{E_i}\}$ ($i=1,2,3$) of a Hamiltonian $\hat H$, namely $\hat{H}\ket{E_{i}}=E_{i}\ket{E_{i}}$, as 
\begin{equation}\label{Psi0}
 \ket{\psi(0)}=\sum_{i=1}^{3}\sqrt{r_{i}}e^{\text{i}\theta_{i}}\ket{E_{i}},
\end{equation}
where $0\le\theta_{i}\le2\pi$,   and    the triad of coefficients $r_{i}$ forms a probability distribution $\{r_{1},r_{2},r_{3}\}$ for which $0\leq r_i\leq 1$ and $\sum_{i=1}^{3}r_{i}=1$. 

We consider initial states whose expansion \eqref{Psi0} involves non-degenerate eigenstates $\ket{E_{i}}$    and    assume that their corresponding eigenvalues $E_{i}$ are ordered according to $E_{1}<E_{2}<E_{3}$. Thus,  the evolved state, $\ket{\psi(t)}=e^{-\text{i}\hat Ht/\hbar}\ket{\psi(0)}$, is given explicitly by
\begin{equation}\label{estado}
\ket{\psi(t)}=\sum_{i=1}^{3}\sqrt{r_{i}}e^{\text{i}\theta_{i}}e^{-\text{i}E_{i}t/\hbar}\ket{E_{i}},
\end{equation}
and the overlap $\braket{\psi(0)}{\psi(t)}$,  which  measures how distinguishable   the state $\ket{\psi(t)}$ is  from the initial one $\ket{\psi(0)}$, reads $ \braket{\psi(0)}{\psi(t)}=\sum_{i=1}^{3}r_{i}e^{-\text{i}E_{i}t/\hbar}$. The system thus attains an orthogonal (distinguishable) state at time $t=\tau$ whenever 
\begin{equation}\label{OverlapGral}
\braket{\psi(0)}{\psi(\tau)}=\sum_{i}r_i e^{-\text{i}E_{i}\tau/\hbar}=0.
\end{equation}
Along with the normalization condition, we have two equations (the real and   imaginary parts of \eqref{OverlapGral}) and four unknowns: $r_1,r_2,r_3$ and $\tau$. In what follows, we identify the families of triads $\{r_i\}$ that solve Equation (\ref{OverlapGral}) in terms of the orthogonality time $\tau$ and the frequencies $\omega_{ij}\equiv(E_{i}-E_{j})/\hbar$. 

\subsection{Families of Allowed Triads $\{r_i\}$}\label{distrib}
{Family  I: Case with $\omega_{ij}\tau= n\pi$ for some $i>j$.} 

In this case,  the coefficients $r_{i}$s that solve Equation (\ref{OverlapGral}) and comply with $\sum_ir_i=1$ satisfy
\begin{subequations}\label{sol2}
\begin{align}
1+r_i[(-1)^n-1]+r_k(\cos\omega_{jk}\tau-1)=0,\label{5a}\\
r_{k}\sin\omega_{jk}\tau=0,\label{5b}
\end{align}
\end{subequations}
with $k\in\{1,2,3\}$ such that $k\neq i,j$. 

From Equation (\ref{5b}), we see that this family of solutions naturally splits into two subfamilies, depending on whether $r_k=0$ or $\sin\omega_{jk}\tau=0$ \cite{BatlePRA2005,ChauPRA2010}. In the first case ($r_k=0$), Equation (\ref{5a}) admits only odd values of $n$,   and    consequently $r_i=r_j=1/2$. This first subfamily thus gives rise to pure states corresponding to the extensively studied case of an effective 
two-level (qubit) system in an equally-weighted superposition. Therefore, from now on, we   refer to this family as Family  I-qubit. Its elements, obtained by varying the three possible values of $k$, read explicitly: 
\begin{equation}\label{FamIq}
\text{I-qubit}=\biggl\{\Bigl\{\frac{1}{2},\frac{1}{2},0\Bigr\},\Bigl\{\frac{1}{2},0,\frac{1}{2}\Bigr\},\Bigl\{0,\frac{1}{2},\frac{1}{2}\Bigr\}\biggr\}. 
\end{equation}

The second subfamily corresponds to the situation for which, in addition to $\omega_{ij}\tau=n\pi$, we have $\sin \omega_{jk}\tau=0$ and $r_{k}\neq0$, so that $\vert\omega_{jk}\vert\tau= m\pi$ with $m$ a positive integer. An element of this subfamily thus reaches an orthogonal state at
\begin{equation}\label{Ib}
 \tau=\frac{n\pi}{\omega_{ij}}=\frac{m\pi}{|\omega_{jk}|}, \quad n,m=1,2,\dots,
\end{equation}
from which it follows that the separations between energy levels are related by $\omega_{ij}/|\omega_{jk}|=n/m$.
According to Equation \eqref{sol2}, we have that odd values of $n$ give the triads: $r_{j}=r_{i}+r_{k}=1/2$ for odd values of $m$, and $r_{i}=r_{j}+r_{k}=1/2$ for even values of $m$;
 while even values of $n$ and odd values of $m$ give $r_{k}=r_{i}+r_{j}=1/2$ (no solution exists for $n$ \emph{and} $m$ even). All the elements of the second subfamily, namely
\begin{equation}\label{FamIb}
\text{I-b}=\biggl\{\Bigl\{\frac{1}{2},r,\frac{1}{2}-r\Bigr\},\Bigl\{r,\frac{1}{2},\frac{1}{2}-r\Bigr\},
\Bigl\{r,\frac{1}{2}-r,\frac{1}{2}\Bigr\}\biggr\},
\end{equation}
with $0<r<1/2$, can be obtained from all the possible sets of indices $\{i,j,k\}$ with $i>j$. 

{Family  II: Case with $\omega_{ij}\tau\neq n\pi$ for all pairs $(i,j)$.} 

In this case,  the orthogonality condition (\ref{OverlapGral}), together with the normalization constraint $\sum_ir_i=1$, implies that the solution coefficients $\{r_i\}$ are of the form
\begin{equation}\label{sol}
r_{i}=\frac{\sin\omega_{jk}\tau}{\sin\omega_{31}\tau+
\sin\omega_{12}\tau+\sin\omega_{23}\tau}, 
\end{equation}
provided the indices $(i,j,k)$ are taken in a cyclic permutation of $(1,2,3)$,   and    $\sin\omega_{31}\tau+
\sin\omega_{12}\tau+\sin\omega_{23}\tau\neq 0$; when this quantity equals zero no solution for $\langle\psi(0)|\psi(\tau)\rangle=0$ and $\sum_ir_i=1$ exists. Consequently, the largest family of triads $\{r_{i}\}$ with non-vanishing elements that solve Equation \eqref{OverlapGral}
is $\text{II}=\{r_{i}\}$, with $r_i$ given by those coefficients of the form (\ref{sol}) that comply with the additional restriction $0< r_i<1$ (the solution $r_i=0$ for some $i$ is excluded since it is already contained in Family  I, whereas the solution $r_i=1$ for some $i$ is ruled out since it corresponds to a stationary state that never reaches orthogonality). Notice that the decomposition $\omega_{31}=\omega_{32}+\omega_{21}$ implies that the triad $\{r_{i}\}$ can be written in terms of $\omega_{21}$, $\omega_{32}$ and $\tau$ only. 

Equations (\ref{FamIq}), (\ref{FamIb}) and (\ref{sol}) determine all the coefficients $\{r_i\}$ that guarantee the evolution of the corresponding initial qutrit (\ref{Psi0}) to an orthogonal state. In the first two cases (Families I-qubit and I-b),
  the coefficients do not depend on the specific Hamiltonian; only the orthogonality time is determined by the energy-level spacing. In Family  II, in contrast, the coefficients, the orthogonality time and the energy-level separations are related in a more complex way. In particular, for a fixed Hamiltonian, in order to determine $\tau$, the set $\{r_i\}$ must be specified,   and,    conversely, to 
determine the latter, the value of $\tau$ must be given. The explicit relation \mbox{(\ref{sol})} may find practical applications
 in determining, for example, which initial state should be prepared for it to become distinguishable at a desired $\tau$ given a certain generator $\hat H$.

\subsection{The Solution Diagram}
\begin{figure}
 \includegraphics[width=\columnwidth]{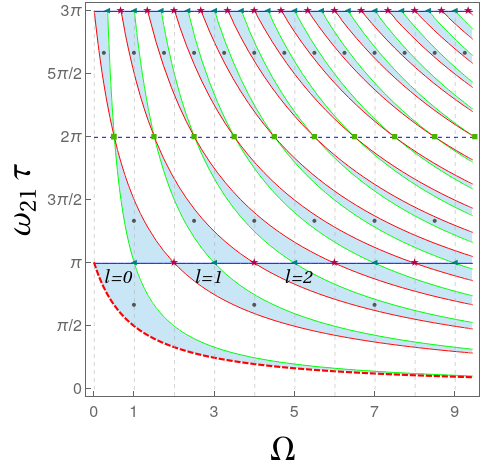}
 \caption{Diagram $\omega_{21}\tau$ vs. $\Omega$, with $\Omega=\omega_{32}/\omega_{21}$. The coefficients $\{r_1,r_2,r_3\}$ giving rise to states \eqref{Psi0} that reach an orthogonal state at time $\tau$  are represented by points in the diagram for the corresponding values of $\omega_{21}\tau$ and $\Omega$. Blue-shaded regions represent $\{r_i\}$ of Family  II, satisfying Equation \eqref{sol}. These regions are bordered by the solutions $\{r_i\}$ pertaining to Family  I as follows. For Subfamily  I-qubit: $\bigl\{\frac{1}{2},\frac{1}{2},0\bigr\}$ is identified with solid-blue lines, $\bigl\{0,\frac{1}{2},\frac{1}{2}\bigr\}$ with solid-green lines   and    $\bigl\{\frac{1}{2},0,\frac{1}{2}\bigr\}$ with red lines. For   Subfamily I-b: $\bigl\{\frac{1}{2},r,\frac{1}{2}-r\bigr\}$ is identified with a star, $\bigl\{r,\frac{1}{2},\frac{1}{2}-r\bigr\}$ with a left-triangle   and    $\bigl\{r,\frac{1}{2}-r,\frac{1}{2}\bigr\}$ with a square. The red-dashed curve indicates the global lower bound for the orthogonality time.}
 \label{Solutions}
\end{figure}

Imposing the condition $0< r_{i}< 1$ on the solutions in Equation \eqref{sol} clearly restricts the possible values of the frequencies $\omega_{ij}$ and   the orthogonality time $\tau$. To get insight into the set of values allowed by this restriction, we identify the elements
 that pertain to Family  II in the phase diagram corresponding to the space determined by the dimensionless orthogonality time $\omega_{21}\tau$ and the ratio $\Omega=\omega_{32}/\omega_{21}$. This diagram is shown in \mbox{Figure \ref{Solutions}},   and    the permitted values---numerically calculated consistently with Equation \eqref{sol} and   the condition $0< r_{i}< 1$---form the `zebra stripe'-like pattern depicted by the blue shaded regions, \emph{excluding} the borders. Thus, each point in the blue region represents a triad $\{r_1,r_2,r_3\}$ that satisfies Equation \eqref{sol} for the corresponding values of $\omega_{21}\tau$ and $\Omega$. Likewise, the points in the borders of the zebra stripes represent triads that pertain to Family  I, as   shown in what follows. 


According to the results in Section \ref{distrib}, for the $\{r_i\}$s of Family  I-qubit,  we have:
\begin{itemize}
\item$\bigl\{\frac{1}{2},\frac{1}{2},0\bigr\}$ corresponds to 
\begin{equation}\label{blue}
\omega_{21}\tau=(2l+1)\pi,\quad l=0,1,\dots, 
\end{equation}
and is represented by the horizontal solid-blue lines in Figure \ref{Solutions}. 

\item$\bigl\{\frac{1}{2},0,\frac{1}{2}\bigr\}$ corresponds to $\omega_{31}\tau=\pi, 3\pi, 5\pi, \ldots$. By writing $\omega_{31}=\omega_{21}(1+\Omega)$, this is equivalently expressed as
\begin{equation} \label{red}
\omega_{21}\tau=\frac{(2l+1)\pi}{1+\Omega}, \quad l=0,1,\dots,
\end{equation}
represented by the red curves.

\item$\bigl\{0,\frac{1}{2},\frac{1}{2}\bigr\}$ corresponds to $\omega_{32}\tau=\pi, 3\pi, 5\pi, \ldots$. With $\omega_{32}=\Omega\omega_{21}$, this amounts to
\begin{equation} \label{green}
\omega_{21}\tau=\frac{(2l+1)\pi}{\Omega}, \quad l=0,1,\dots,
\end{equation}
represented by solid-green curves.
\end{itemize}

For the triads in Family  I-b, an analysis of the cases described below Equation (\ref{Ib}) shows that:
\begin{itemize}
 \renewcommand{\labelitemi}{\tiny$\blacksquare$}
\item$\Bigl\{\frac{1}{2},r,\frac{1}{2}-r\Bigr\}$ corresponds to
\begin{equation}
\omega_{21}\tau=(2l+1)\pi \;\;\;\textrm{and}\;\;\;\Omega=\frac{2(l'+1)}{2l+1},
\end{equation}
with $l,l'=0,1\dots$. Consequently,  the points representing these triads
 (marked with a star symbol) are found in the intersections of the solid-blue lines with the lines (not shown in Figure \ref{Solutions}) $\Omega=2(l'+1)/(2l+1)$.

\item $\Bigl\{r,\frac{1}{2},\frac{1}{2}-r\Bigr\}$ corresponds to 
\begin{equation}
\omega_{21}\tau=(2l+1)\pi \;\;\;\textrm{and}\;\;\;\Omega=\frac{2l'+1}{2l+1},
\end{equation}
with $l,l'=0,1\dots$. 
The triads are therefore located at the intersections (marked with the left-triangle symbol) of the solid-blue lines with the lines (not shown) $\Omega=(2l'+1)/(2l+1)$.

\item $\Bigl\{r,\frac{1}{2}-r,\frac{1}{2}\Bigr\}$ corresponds to 
\begin{equation}
\omega_{21}\tau=2(l+1)\pi \;\;\;\textrm{and}\;\;\;\Omega=\frac{2l'+1}{2(l+1)},
\end{equation}
with $l,l'=0,1\dots$. The set
 of coefficients are thus located at the intersections (marked with square symbol) of the dashed-blue lines with the lines (not shown) $\Omega=(2l'+1)/2(l+1)$.
\end{itemize}

The above results confirm that the borders of the blue shaded regions do indeed represent solutions in Family  I. The interior of the zebra stripes, as stated above, contains those that pertain to Family  II, which can be located arbitrarily near to the borders. This can be verified as follows. For simplicity, we consider the zebra stripes (labeled as $l=0,1,\ldots$) contained in the region $\mathcal{R}_{1}=(0,\infty)\times(0,\pi)$ of the diagram $\omega_{21}\tau$ vs. $\Omega$. In $\mathcal{R}_{1}$, we have that $\sin\omega_{21}\tau> 0$, which, together with the condition $r_3> 0$, implies that 
the denominator in Equation (\ref{sol}) is a negative quantity; consequently, $r_1> 0$ gives $\sin\omega_{32}\tau> 0$, which leads to the conditions
\begin{subequations}\label{conds}
\begin{align}
0&<\omega_{21}\tau<\pi,\\
2l\pi&<\omega_{32}\tau<(2l+1)\pi,
\end{align}
for $l=0,1,\dots$. By use of these inequalities and the relation $\omega_{31}=\omega_{32}+\omega_{21}$, we have $2l\pi<\omega_{31}\tau<2(l+1)\pi$. However, the condition $r_2> 0$ implies that $\sin \omega_{31}\tau < 0$,   and    consequently the bounds of $\omega_{31}\tau$ are tightened to
\begin{equation}
(2l+1)\pi<\omega_{31}\tau<2(l+1)\pi. 
\end{equation}
\end{subequations}

Now, by writing $\omega_{32}=\Omega\omega_{21}$    and    $\omega_{31}=\omega_{21}(1+\Omega)$, the inequalities \eqref{conds} are satisfied for each $l$ if
\begin{equation}\label{cotasAp}
 \max{\left\{\frac{2l\pi}{\Omega}, \frac{(2l+1)\pi}{1+\Omega}\right\}}<\omega_{21}\tau < \min\biggl\{\pi,
\frac{(2l+1)\pi}{\Omega}, \frac{2(l+1)\pi}{1+\Omega}\biggr\},
\end{equation}
obtaining finally that the $l$th zebra stripe in $\mathcal{R}_{1}$ is correspondingly delimited by the boundaries expressed in the inequalities 
\begin{equation}\label{cotasgral}
 \frac{(2l+1)\pi}{1+\Omega}< \omega_{21}\tau<
 \begin{cases}
 \pi & 2l\le\Omega\leq 2l+1,\\
 \dfrac{(2l+1)\pi}{\Omega} & \Omega\geq 2l+1.
 \end{cases}
\end{equation}
The lower and upper bounds in this expression correspond, respectively, to the red and   green (or blue) borders seen above. Therefore, since $\omega_{21}\tau$ can be arbitrarily close to the bounds, there is a point inside the zebra stripes that approximates to the borders as much as desired. Physically, this means that there is a triad $\{r_i\}$ for which the orthogonality time can be arbitrarily close to the limiting values in Equation (\ref{cotasgral}).

\subsection{Orthogonality Time}
From the lower bound of Equation (\ref{cotasgral}) , we get the minimal orthogonality time for the $l$th zebra stripe:
\begin{equation}\label{borders}
\tau_{<}(l)=\frac{(2l+1)\pi}{1+\Omega}\frac{1}{\omega_{21}}=\frac{(2l+1)\pi}{\omega_{31}}.
\end{equation}
Taking $l=0$ in Equation (\ref{borders}),  we get 
\begin{equation}\label{taumin}
\tau_{\min}\equiv\min_{\{l\}}[\tau_{<}(l)]=\tau_{<}(0)=\frac{\pi}{\omega_{31}},
\end{equation}
thus $\tau_{<}(0)$---represented by the red-dashed curve in Figure \ref{Solutions}---
 stands for the \emph{global minimal} possible amount of time required to reach orthogonality. Since $\tau_{\min}\sim \omega_{31}^{-1}$, orthogonality can be reached more quickly  as the separation between the extreme levels 
increases. Recall, however, that this minimal time is not attainable for states in Family  II, lying inside the zebra stripe;  however, as discussed above, it is always possible to find a point inside the blue-shaded area corresponding to a three-level state that attains orthogonality at a time arbitrarily close to $\tau_{\min}$. 

The two upper bounds in Equation (\ref{cotasgral}) coincide whenever $\Omega=2l+1$.
For $l=0$,  this occurs when the separation between levels coincide, so that $\omega_{32}=\omega_{21}=\omega$    and    $\omega_{31}=2\omega$. This equally-spaced case is indicated with the vertical dashed line $\Omega=1$ in Figure \ref{Solutions}    and    corresponds, according to Equation \eqref{sol}, to 
\begin{subequations}\label{solEqS}
\begin{align}
r_{2}&=\frac{\cos\omega\tau}{\cos\omega\tau-1}, \label{11beq} \\
r_{3}&=r_1=\frac{1}{2(1-\cos\omega\tau)}.\label{11ceq}
\end{align}
\end{subequations}

As a particular example of this equally-spaced energy levels case, we find the $\tau$s for the equally-probable superposition, corresponding to $r_1=r_2=r_3=1/3$ (represented in Figure \ref{Solutions} by dark dots at the `center' of each zebra stripe). From Equation \eqref{solEqS}, we have that $\omega\tau$ must satisfy $\cos\omega\tau=-1/2$, which leads to
\begin{equation}
\tau=\frac{2\pi}{3\omega},\, \frac{4\pi}{3\omega},\, \frac{8\pi}{3\omega},\ldots.
\end{equation}
The first of these values, $\tau_1=2\pi/3\omega$ (corresponding to the dot inside the $l=0$ zebra stripe), determines the time at which $\ket{\psi(t)}=\frac{1}{\sqrt{3}}\sum_{i}e^{\text{i}\theta_{i}}e^{-\text{i}E_{i}t/\hbar}\ket{E_{i}}$ becomes distinguishable from the initial state $\ket{\psi(0)}$ for the first time, while the second, $\tau_2=2\tau_1$ (second point on the vertical line $\Omega=1$), gives the time at which a second distinguishable state, $\ket{\psi(2\tau_1)}$, orthogonal to $\ket{\psi(0)}$ \emph{and}   $\ket{\psi(\tau_1)}$, is reached \cite{ValdesJPhysA2020}. This is a particular example (for $\mathcal{N}=3$) of the previously studied case of equally-weighted superpositions of $\mathcal{N}$ non-degenerate and equally-spaced states \cite{ValdesJPhysA2020}. 
Furthermore, the regular distribution of the dark dots in the middle of each zebra stripe is a consequence of a more general property, namely, that each distribution corresponding to a given $\tau$ and $\Omega$ in stripe $l=0$  periodically appears in all other stripes at the same $\tau$    and      $\Omega_l=\Omega+2\pi l/\tau$. 

In Figure \ref{Solutions}, we observe that, for $0<\Omega\leq1$, there exist triads of Family  II that give rise to states $\ket{\psi(0)}$ and $\ket{\psi(\tau)}$ that are mutually orthogonal for \emph{any} $\tau$ in the interval $\tau_{\min}=\pi/\omega_{31}< \tau < \pi/\omega_{21}$.
 For $\Omega>1$ in the region $\mathcal{R}_{1}$, the allowed values of the orthogonality time are grouped into bands, whose number increases, decreasing their width, as $\Omega$ acquires higher values. In any case, it should be noted that \emph{at least} one solution exists for \emph{all} $\Omega$, meaning that a  three-level system can be made to reach an orthogonal state in a finite time $\tau<\pi/\omega_{21}$, provided the expansion coefficients $\{r_i\}$,  which  specify the initial state preparation \eqref{Psi0}, are adequately chosen.

\begin{figure}
\includegraphics[width=\columnwidth, trim=0 70 0 10]{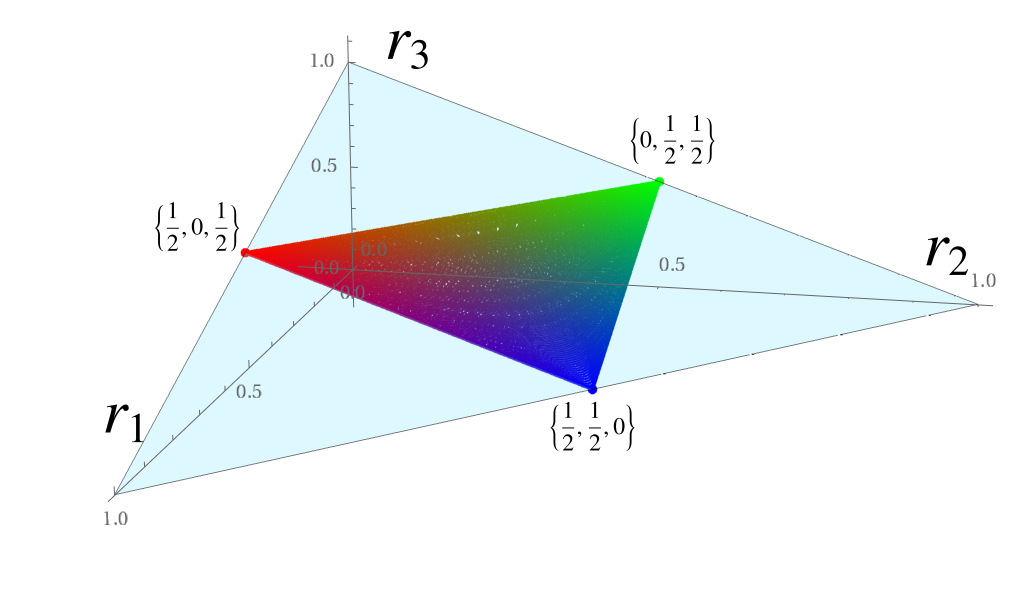}
\caption{The 2-simplex $\Delta^{2}$ of $\mathbb{R}^{3}$ (light-blue shaded plane), defined by the set of points $(r_{1},r_{2},r_{3})$ satisfying $r_i\geq0$ and $\sum_{i}r_{i}=1$. The 2-simplex $\delta^{2}$ (colored central triangle) contains the subset of points of $\Delta^{2}$ that define the coefficients in the energy-expansion of initial states $\ket{\psi(0)}$ that evolve towards a distinguishable state at time $\tau$. The vertices of $\delta^{2}$ correspond to elements of Family  I-qubit, its edges to elements of Family  I-b   and    its interior to elements of Family  II. The colors in $\delta^{2}$ identify the triads according to a RGB map-code defined by its vertices.}
 \label{fig:rSpace}
\end{figure}

Before ending this section,   and    in order to gain more insight into the geometric representation of the allowed $\{r_{i}\}$s, we now focus on the space $(r_1,r_2,r_3)$ and identify in it the triads for which mutually orthogonal states, $\ket{\psi(\tau)}$ and $\ket{\psi(0)}$, exist. Such points form a subset $\delta^{2}$ of the \emph{standard 2-simplex} $\Delta^{2}$ in $\mathbb{R}^{3}$, the latter defined by the points $(r_1,r_2,r_3)$ that satisfy $r_i\geq 0$ and $\sum_{i}r_{i}=1$    and    represented by the light-blue shaded face in Figure \ref{fig:rSpace}. $\delta^{2}$, which is in itself a 2-simplex, is depicted as the colored triangle. Its vertices and edges (its boundary) correspond, respectively, to the elements of the Families  I-qubit and I-b,   and    its interior is filled with by those $\{r_i\}$ in Family  II. As can be seen in  the figure, these latter points do satisfy $0<r_i<1/2$, meaning, in particular, that for highly unbalanced superpositions of energy eigenstates the orthogonality condition cannot be met.

\section{\label{sect:QSL}The Quantum Speed Limit}
It is a well-established fact that, for given energetic resources, the orthogonality time of an initial state cannot be less than the (minimal) time imposed by the so-called \emph{quantum speed limit}. The QSL establishes a natural and intrinsic time-scale of the system's dynamics    and    becomes relevant when characterizing the evolution rate of the system. It is therefore the central quantity of this section, aimed at determining the quantum speed limit of the states associated to the expansion coefficients that belong to the Families I and II.

The fundamental limits on the speed at which a pure quantum state evolves towards an orthogonal one under a (time-independent) Hamiltonian have been advanced in the form of lower bounds, either in terms of the relative mean energy $\mathcal{E}$ (as measured from the lowest eigenvalue that contributes to \eqref{Psi0}, say $E_j$)  \cite{Mandelstam1991}, 
\begin{subequations}\label{BOunds}
\begin{equation}\label{ML}
\tau\geq\frac{\pi\hbar}{2\mathcal{E}}, \quad \mathcal{E}=\mean{(\op{H}-E_{j})},
\end{equation}
or in terms of the energy dispersion \cite{MargolusPhysicaD1998} $\sigma_{\op{H}}=\sqrt{\mean{\op{H}^{2}}-\mean{\op{H}}^{2}}$, 
\begin{equation}\label{MT}
\tau\geq\frac{\pi\hbar}{2\sigma_{\op{H}}},
\end{equation}
with $\mean{\op{H}^n}=\sum_{i}r_{i}E^n_{i}$. Equations \eqref{ML} and \eqref{MT} are, respectively, the celebrated Margolus--Levitin (ML) and Mandestam-Tamm (MT) bounds. In \cite{LevitinPRL2009}, Levitin and Toffoli considered the unified bound \cite{LevitinPRL2009}
\begin{equation}
\label{Bound}
\tau\geq\tau_{\text{\textrm{qsl}}}\equiv\max\left\{\frac{\pi\hbar}{2\mathcal{E}},\frac{\pi\hbar}{2\sigma_{\op{H}}}\right\},
\end{equation}
\end{subequations}
with $\tau_{\text{\textrm{qsl}}}$ denoting the quantum speed limit. They proved that only for states in an equally-weighted superposition of two energy eigenstates both bounds (ML and MT) coincide \emph{and} the quantum speed limit is attained, whereas for superpositions of three energy eigenstates the unified bound is tight, so $\tau$ may be arbitrarily close, although not equal, to $\tau_{\text{qsl}}$. This means that for the triads $\{r_i\}$ in Family  I-qubit $\tau_{\text{qsl}}=\tau$, whereas for those in the Families I-b and II it holds that $\tau_{\text{qsl}}<\tau$. Our aim is thus to discern which of the two bounds, ML or MT, determines the quantum speed limit of the states \eqref{Psi0} specified by the elements of these latter families. We   do so by analyzing the parameter
\begin{equation}
 \alpha=\frac{\sigma_{\op{H}}}{\mathcal{E}},
\end{equation}
which evidently determines $\tau_{\text{qsl}}$ as the Mandelstam--Tamm bound whenever $\alpha<1$,        as the Margolus--Levitin bound provided $\alpha>1$   and    by either of them  (being equal) for $\alpha=1$. 

In terms of the transition frequencies $\omega_{ij}$ and the coefficients $r_i$, $\mathcal{E}$ and $\sigma_{\op{H}}$ are explicitly given by 
\begin{subequations}\label{means}
\begin{align}
\mathcal{E}&=\mean{\op{H}}-E_{j}=\hbar\omega_{ij}r_i+\hbar\omega_{kj}r_k,\label{Esigma} \\
\sigma_{\op{H}}&=\sqrt{\frac{\hbar^2}{ 2}\sum_{i,k=1}^{3}r_ir_k\omega^2_{ik}}.\label{sigma}
\end{align}
\end{subequations}
For the triads in Family  I-qubit, Equation \eqref{means} reduces to $\mathcal{E}=\sigma_{\op{H}}=\hbar\omega_{ij}/2$; therefore, $\alpha=1$,     bounds \eqref{ML} and \eqref{MT} coincide,   and    it is verified that $\tau=\tau_\text{qsl}$, with $\tau_\text{qsl}=\pi/\omega_{ij}$. Notice that,  among the three members of Family  I-qubit, the one with the lowest quantum speed limit is $\bigl\{\frac{1}{2},0,\frac{1}{2}\bigr\}$ (represented by the red vertex of $\delta^2$ in \mbox{Figure \ref{fig:rSpace})}, which gives rise to the fastest qubit $\ket{\psi(0)}=\frac{1}{\sqrt{2}}\Bigl(e^{\text{i}\theta_{1}}\ket{E_{1}}+e^{\text{i}\theta_{3}}\ket{E_{3}}\Bigr)$  that reaches a distinguishable state at $\tau$ given precisely by Equation \eqref{taumin}.

\begin{figure}
 \includegraphics[width=\columnwidth]{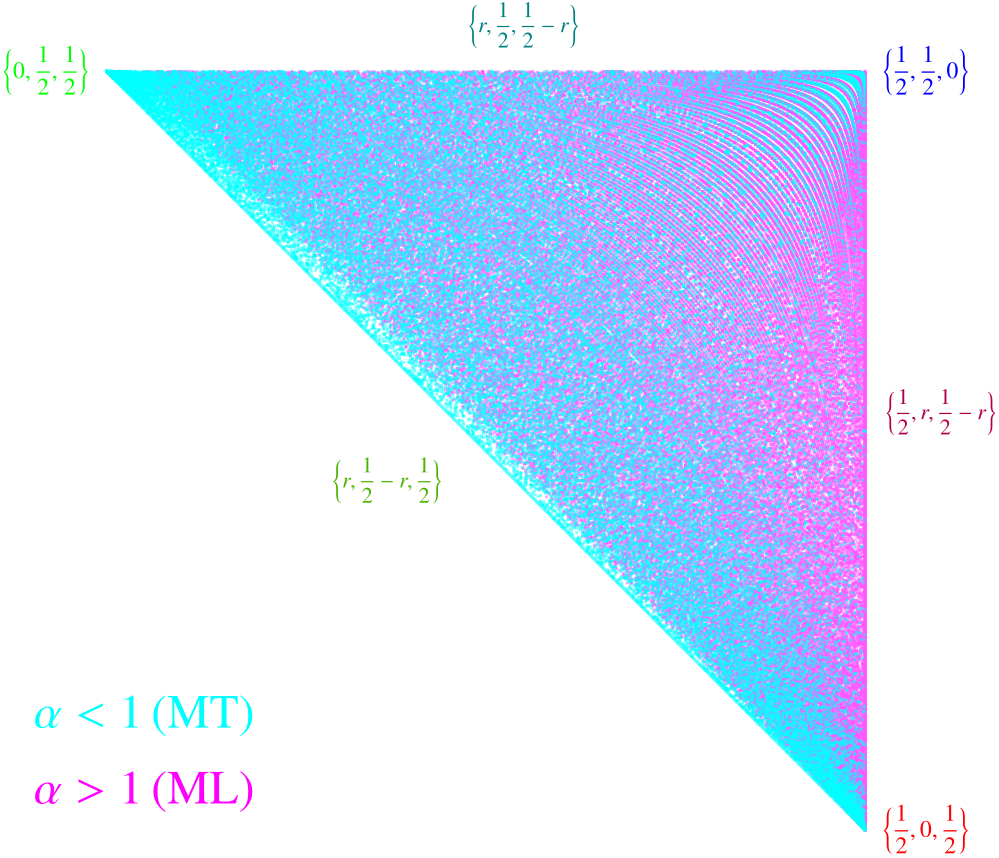}
 \caption{Map of the quantum speed limit (of the states associated to the triads $\{r_i\}$) in the 2-simplex $\delta^{2}$. For different values of $\Omega$, the simplex is colored according to the range of $\alpha$ (Equation \eqref{alpha}) as follows: the Mandelstam--Tamm subset $\delta^{2}_\text{MT}$ ($\alpha<1$, cyan) and the Margolus--Levitin one $\delta^{2}_\text{ML}$ ($\alpha>1$, magenta).}
 \label{fig:Simplex}
\end{figure}

When all three coefficients $r_i$ are nonzero, we can rewrite Equation (\ref{means}) as follows
\begin{subequations}\label{Esig}
\begin{align}
\mathcal{E}&=\mean{\op{H}}-E_{1}=\hbar\omega_{21}[r_2+r_3(1+\Omega)],\label{GMean}\\
\sigma_{\op{H}}&=\hbar\omega_{21}\sqrt{r_1r_2+r_1r_3(1+\Omega)^2+r_2r_3\Omega^2},\label{Esigma}
\end{align}
\end{subequations}
from which $\alpha$ can be expressed as
\begin{equation}\label{alpha}
 \alpha=\sqrt{\frac{r_{2}+r_{3}\left(1+\Omega\right)^{2}}{\big[r_{2}+r_{3}(1+\Omega)\big]^{2}}-1}.
\end{equation}
This allows the realization of a quantitative analysis of the quantum speed limit of states given rise by Families I and II.
For an arbitrary fixed $\{r_{i}\}$, $\alpha$ depends \emph{explicitly} on the energy separations only through the ratio $\Omega=\omega_{32}/\omega_{21}$. However, for the $\{r_{i}\}$ in Family  II, the \emph{detailed} dependence of $\alpha$ on $\Omega$ becomes more intricate due to Equation \eqref{sol}. In Figure \ref{fig:Simplex}, we show a large sample of points of the 2-simplex $\delta^{2}$ that correspond to coefficients for which $\alpha$, computed from \eqref{alpha}, is larger than 1 (the Margolus--Levitin subset $\delta^{2}_\text{ML}$, colored in magenta),   and    similarly coefficients for which $\alpha<1$ (the Mandelstam--Tamm subset $\delta^{2}_\text{MT}$, colored in cyan), for different values of $\Omega$ in the region $(0,3\pi^{2}]\times(0,18\pi]$ of the space $\omega_{21}\tau$--$\Omega$. The figure reveals the complex spreading of both subsets within the orthogonality simplex $\delta^{2}$, induced by the quantum speed limit; notably, there is no sharp separation between the elements of $\delta^{2}_\text{MT}$ and those of $\delta^{2}_\text{ML}$. They appear rather mixed, although exhibiting a marked accumulation of cyan points around the edge $\bigl\{r,\frac{1}{2}-r,\frac{1}{2}\bigr\}$ and the vertex $\bigl\{0,\frac{1}{2},\frac{1}{2}\bigr\}$. These become diluted towards the opposite vertex $\bigl\{\frac{1}{2},\frac{1}{2},0\bigr\}$, whereas the magenta points are more dense near the vertices $\bigl\{\frac{1}{2},0,\frac{1}{2}\bigr\}$ and $\bigl\{\frac{1}{2},\frac{1}{2},0\bigr\}$.

The case of the triads in Family  I-b, represented by the edges of $\delta^{2}$ in Figure \ref{fig:Simplex}, allows for a direct and simple analytical analysis for determining the corresponding range of $\alpha$. With this purpose,  it is convenient to analyze separately the three possible sets $\{r_i=1/2=r_j+r_k\}$ as follows:
\begin{itemize}
\renewcommand{\labelitemi}{\tiny$\blacksquare$}
\item $\bigl\{\frac{1}{2},r,\frac{1}{2}-r\bigr\}$, $0< r<\frac{1}{2}$.
These points form the edge that goes from the vertex $\bigl\{\frac{1}{2},0,\frac{1}{2}\bigr\}$ to $\bigl\{\frac{1}{2},\frac{1}{2},0\bigr\}$ of the 2-simplex $\delta^{2}$ in Figures \ref{fig:rSpace} and \ref{fig:Simplex},   and   they   correspond to
\begin{equation}
\alpha=\sqrt{1+\frac{r(1-2r)\Omega^2}{\big[r+(\frac{1}{2}-r)(1+\Omega)\big]^{2}}}>1\quad \textrm{for all}\;\: \Omega,
\end{equation}
where the inequality follows from the fact that $0<1-2r<1$.

\item $\bigl\{r,\frac{1}{2}-r,\frac{1}{2}\bigr\}$, $0< r<\frac{1}{2}$. These points form the edge that goes from the vertex $\bigl\{\frac{1}{2},0,\frac{1}{2}\bigr\}$ to $\bigl\{0,\frac{1}{2},\frac{1}{2}\bigr\}$ in Figures \ref{fig:rSpace} and \ref{fig:Simplex},   and    give
\begin{equation}
\alpha=\sqrt{1-\frac{(1-2r)(1-r+\Omega)}{\big[1-r+\frac{1}{2}\Omega\big]^{2}}}<1\quad \textrm{for all}\;\: \Omega.
\end{equation}

\item $\bigl\{r,\frac{1}{2},\frac{1}{2}-r\bigr\}$, $0< r<\frac{1}{2}$. This set of points lies along the edge that goes from the vertex $\bigl\{0,\frac{1}{2},\frac{1}{2}\bigr\}$ to $\bigl\{\frac{1}{2},\frac{1}{2},0\bigr\}$ in Figures \ref{fig:rSpace} and \ref{fig:Simplex},   and    their corresponding $\alpha$ reads
\begin{equation}
\alpha=\sqrt{1+\frac{(1-2r)(1+\Omega)\big[r(1+\Omega)-1\big]}{\big[1-r+\Omega\bigl(\frac{1}{2}-r\bigr)\big]^{2}}}.
\end{equation}

Unlike the previous cases, here, the range of $\alpha$ depends on the range of $\Omega$. For $\Omega\le1$ we have $\alpha<1$, whereas for $\Omega>1$ we find three cases: 
\begin{subequations}
\begin{align}
 \alpha>1& \;\;\text{for} \;\; \frac{1}{1+\Omega}<r<\frac{1}{2},\\
 \alpha=1&\;\;\text{for} \;\; r=\frac{1}{1+\Omega},\\
 \alpha<1& \;\;\text{for} \;\; 0<r<\frac{1}{1+\Omega}. 
 \end{align}
\end{subequations} 

In this way, for a given $\Omega$, the edge under consideration is divided into two segments, one cyan and one magenta, separated by the point at $r=(1+\Omega)^{-1}$. Notice that such segments are not appreciated in Figure \ref{fig:Simplex}, since (as explained above) this figure considers a sample of different values of $\Omega$    and    not a single one. 
\end{itemize} 

Finally, Table \ref{T} contains a practical summary of our main results, which completely characterize all qutrits that evolve towards an orthogonal state under a time-independent, otherwise arbitrary, Hamiltonian.

\begin{table*}
\begin{center}
\caption{Complete set of expansion coefficients $\{r_i\}$ that satisfy the orthogonality condition, the corresponding orthogonality time $\tau$ (in terms of the energy-separation levels of the Hamiltonian), and ratio $\alpha$ that determines whether the quantum speed limit is given by Mandelstam-Tamm bound ($\alpha<1$) or the Margolus-Levitin bound ($\alpha>1$). We have denoted $D={\sin\omega_{23}\tau+\sin\omega_{31}\tau+\sin\omega_{12}\tau}$.}\label{T}
\begin{tabular}{ c c c c c |c| c }
\hline
\bf{Family}&$\boldsymbol{r_1}$ & $\boldsymbol{r_2}$ & $\boldsymbol{r_3}$ & & $\boldsymbol{\tau}$ & $\boldsymbol{\alpha}$\\
\hline 
\\[-1em]
 I-qubit&0&$\frac{1}{2}$   & $\frac{1}{2}$ && $\frac{n\pi}{\omega_{32}}$, $n$ odd &1\\
\\[-1em]
\hline
\\[-1em]
I-qubit&$\frac{1}{2}$ &  0 & $\frac{1}{2}$ && $\frac{n\pi}{\omega_{31}}$, $n$ odd &1\\ 
\\[-1em]
\hline
\\[-1em]
I-qubit&$\frac{1}{2}$ & $\frac{1}{2}$   & 0 && $\frac{n\pi}{\omega_{21}}$, $n$ odd &1\\
\\[-1em]
\hline \hline
\\[-1em]
I-b&$\frac{1}{2}$ &  $r$ & $\frac{1}{2}-r$ & $(0<r<\frac{1}{2})$ & $\frac{n\pi}{\omega_{21}}$ with $\frac{\omega_{32}}{\omega_{21}}=\frac{m}{n}$, $n$ odd, $m$ even &$>1$\\ 
\\[-1em]
\hline
\\[-1em]
I-b&$r$ &  $\frac{1}{2}-r$ & $\frac{1}{2}$ &  $(0<r<\frac{1}{2})$ & $\frac{n\pi}{\omega_{21}}$ with $\frac{\omega_{32}}{\omega_{21}}=\frac{m}{n}$, $n$ even, $m$ odd &$<1$ \\ 
\\[-1em]
\hline
\\[-1em]
I-b&$r$ &  $\frac{1}{2}$ & $\frac{1}{2}-r$ & $(0<r<\frac{1}{2})$& $\frac{n\pi}{\omega_{21}}$ with $\frac{\omega_{32}}{\omega_{21}}=\frac{m}{n}$, $n$ odd, $m$ odd &\begin{tabular}{l} $>1$, for $\frac{1}{1+\frac{\omega_{32}}{\omega_{21}}}<r<\frac{1}{2}$ \\ $=1$, for $r=\frac{1}{1+\frac{\omega_{32}}{\omega_{21}}}$ \\ $<1$, for $0<r<\frac{1}{1+\frac{\omega_{32}}{\omega_{21}}}$ \\ 
\end{tabular} \\ 
\\[-1em]
\hline\hline
\\[-1em]
II&$\frac{\sin\omega_{23}\tau}{D}$ &  $\frac{\sin\omega_{31}\tau}{D}$ & $\frac{\sin\omega_{12}\tau}{D}$ &$(0<r_i<1)$& implicitly defined via Eq. (\ref{sol}) &$\gtreqqless 1$ \\ 
\\[-1em]
\hline
\end{tabular}
\end{center}
\end{table*}

\section{\label{sect:Conclusions}Summary and Final Remarks}
In this paper,  we completely determine  and organize  the set of coefficients $\{r_{i}\}$ that provide the energy distribution and give rise to initial states (\ref{Psi0}) that reach an orthogonal state in a finite time $\tau$, when evolving under an arbitrary time-independent Hamiltonian. A geometric organization of such coefficients is established, both in the solution diagram in the space $\omega_{21}\tau$-$\Omega$   and    in the 2-simplex in $\mathbb{R}^{3}$. In the first case, the sets $\{r_{i}\}$ are represented by points according to the allowed values of $\tau$ and the corresponding energy-levels separation,   and    characteristic regions whose shape resembles zebra-like stripes emerge. The interior of each of these regions is filled with triads of Family  II ($r_i\neq 0,1/2$ for all $i$), whereas the borders correspond to triads of Family  I: continuous borders represent coefficients associated to qubit states ($r_i=0$ for some $i$),   and    their intersections represent elements of   Subfamily  I-b ($r_i=1/2$ for some $i$, $r_j\neq 0$ for all $j$). In the second geometric representation, the $\{r_i\}$s are organized in the central 2-simplex $ \delta^2$, contained in the 2-simplex $\Delta$ of $\mathbb{R}^{3}$. In this case as well, the elements in Family  II fill the interior (satisfying $0<r_i<1/2)$, whereas elements in Family  I lie along the borders (edges and vertices) of $\delta^2$. 

As is clear from the results in  Figure \ref{Solutions}, the minimum orthogonality time $\tau_\text{min}$ \eqref{taumin} of the qubit corresponding to the triad $R_{13}\equiv\bigl\{\frac{1}{2},0,\frac{1}{2}\bigr\}$ (dotted red line)  bounds from below the orthogonality time of any other qubit or qutrit with the same energetic resources. The other qubits that attain an orthogonal state (representing equally weighted superpositions of eigenstates with energies $E_1$ and $E_2$    and    $E_2$ and $E_3$, respectively)  correspond to the triads $R_{12}\equiv\bigl\{\frac{1}{2},\frac{1}{2},0\bigr\}$ and $R_{23}\equiv\bigl\{0,\frac{1}{2},\frac{1}{2}\bigr\}$   and    have orthogonality times that are bounded according to: $\tau_\text{min}<\tau_{R_{12}}\le\tau_{R_{23}}$ whenever $\omega_{21}\ge\omega_{32}$    and    $\tau_\text{min}<\tau_{R_{23}}\le\tau_{R_{12}}$ provided $\omega_{32}\ge\omega_{21}$ (see Equations \eqref{blue} and \eqref{green} with $l=0$). In the latter case, there exist qutrits whose orthogonality times may acquire any value within the interval $(\tau_\text{min},\tau_{R_{23}})$, whose length diminishes as $\omega_{32}/\omega_{21}$ increases. Such qutrits are represented in the blue shaded region of the $l=0$ zebra stripe in Figure \ref{Solutions}. 

We furthermore analyze  the quantum speed limit of the pure states given rise by the points within the 2-simplex $\delta^{2}$, by constructing a map where the states whose orthogonality time is bounded by the Margolus--Levitin bound are distinguished from those for which the Mandelstam--Tamm bound limits the time evolution towards orthogonality. 

Our investigation constitutes a comprehensive analysis of the exact solutions of the orthogonality condition $\braket{\psi(0)}{\psi(\tau)}=0$ in three-level systems. It reveals a rich underlying geometric structure and hierarchy of the allowed parameters that conform the energy distributions, while allowing for the establishment of limiting values for the orthogonality time in terms of the energy levels spacing. Further, knowledge of the expansion coefficients and their relation with $\tau$ provides central tools to analyze the detailed dynamics of qutrits evolving towards orthogonality    and      explore its relation with other relevant features, such as the amount of entanglement in composite three-level systems, which  strongly depends on the sets $\{r_i\}$. 

We consider that our study contributes   from both theoretical and   practical points of view to the efforts ultimately aimed at establishing the conditions under which a $N$-level state transforms into a distinguishable one under a unitary transformation, irrespective of the peculiarities of the physical system. In this sense,  it should be stressed that, although we   assume  a Hamiltonian evolution, the analysis just performed can be immediately extended to more general continuous unitary transformations $e^{-\text{i}\hat G\gamma}$, with $\hat G$ an appropriate ($\gamma$-independent) generator    and    $\gamma$ the corresponding evolution parameter. Our findings are thus applicable to any pure state (of single or composite systems) that can be expressed as a superposition of any three, non-degenerate, eigenvectors of $\hat G$    and    throw light onto  the $\gamma$-distribution and the \emph{amount of evolution} $\gamma^{\bot}$ required to reach an orthogonal state under the transformation governed by $\hat G$. The results here presented thus acquire relevance in the realm of the dynamics of low-dimensional states,    as well as  in the field of quantum information processing, in which knowing and approximating to the fundamental limits of a quantum dynamical process improves quantum control technologies.
\acknowledgements
This work was supported by UNAM-PAPIIT IN110120 (F.J.S. and A.J.B.), and UNAM-PAPIIT IN113720 (A.V.H).

\end{document}